\documentclass{article}

\usepackage{graphicx}

\def\abstract#1{\vskip 7mm 
        \begin{center}{\large Abstract}\par \smallskip
                \begin{minipage}[c]{12cm}
                        \small #1
                \end{minipage}
        \end{center}
}
\def\title#1{\begin{center}{\Large\bf #1}\end{center}}
\def\author#1{\vskip 5mm \begin{center}{#1}\end{center}}
\def\address#1{\begin{center}{\it #1}\end{center}}

\def\Tr{{\rm Tr}}
\def\tr{{\rm tr}}

\makeatletter
\@ifundefined{lesssim}{}{}
\@ifundefined{gtrsim}{}{}
\def\vereq#1#2{\lower3pt\vbox{\baselineskip1.5pt \lineskip1.5pt
\ialign{$\m@th#1\hfill##\hfil$\crcr#2\crcr\sim\crcr}}}
\makeatother

\begin{document}

\title{%
Induced Gravity in Deconstructed Space 
at Finite Temperature
  \smallskip \\
  {\large --- Self-consistent Einstein Universe  ---}
}
\author{%
  Nahomi Kan\footnote{E-mail:kan@yamaguchi-jc.ac.jp}
  and
  Kiyoshi Shiraishi\footnote{E-mail:shiraish@sci.yamaguchi-u.ac.jp}
}
\address{%
$^1$Yamaguchi Junior College, 1346--2 Daidou, Hofu-shi, Yamaguchi
747--1232, Japan\\
$^2$
Yamaguchi University, 1677--1 Yoshida,
Yamaguchi-shi, Yamaguchi 753--8512, Japan
}

\abstract{
We study self-consistent cosmological solutions for an Einstein Universe
in a graph-based induced gravity model. 
The graph-based field theory has been proposed by the present authors to generalize
dimensional deconstruction. 
In this paper, we consider self-consistent Einstein equations 
for a ``graph theory space''.
Especially, we demonstrate specific results for cycle graphs.
}

\section{Pre-history}

\subsection{Induced Gravity}
Induced Gravity or Emergent Gravity has been studied by many authors.
The idea of induced gravity is,
``Gravity emerges from the quantum effect of matter fields.''
The one-loop effective action can be expressed as the form: 
\begin{equation}
-\frac{1}{2}\int\frac{dt}{t}\sum_i \Tr\exp\left[-(-\nabla^2+M^2_i)t\right].
\end{equation}
In curved D-dimensional spacetime,
the trace part including the D-dimensional Laplacian becomes
\begin{equation}
\Tr\exp\left[-(-\nabla^2)t\right]=\frac{\sqrt{|\det~ g_{\mu\nu}|}}{(4\pi)^{D/2}}t^{-D/2}
(a_0+a_1 t+\cdots), 
\end{equation}
where
the coefficients depend on the background fields. 
In four-dimensional spacetime,
the coefficients are $a_0=1$ and $a_1=R/6$ for a minimal scalar field,
$a_0=-4$ and $a_1=R/3$ for a Dirac field, 
$a_0=3$ and $a_1=-R/2$ for a massive vector field,
where $R$ is the scalar curvature.

In Kaluza-Klein (KK) theories,
inducing Einstein-Hilbert term were also studied \cite{Toms}.
In Dimensional Deconstruction (see the next subsection), 
we also have constructed models of induced gravity
based on a graph \cite{KSPTP}.

\subsection{Dimensional Deconstruction}\label{DD}
Dimensional Deconstruction (DD) \cite{Deconstruction} is equivalent to 
a higher-dimensional theory with discretized extra dimensions
at a low energy scale.
The Lagrangian density for vector fields is
\begin{equation}
\mathcal{L}=-\frac{1}{2g^2}\sum_{k=1}^N 
{\rm \tr}F_{\mu\nu k}^2+\sum_{k=1}^N 
{\rm tr}\left|D_\mu U_{k,k+1}\right|^2,
\end{equation}  
where $F_{k}^{\mu\nu}=\partial^\mu A^\nu_k 
- \partial^\nu A^\mu_k-i[A^\mu_k,A^\nu_k]%
$ 
is the field strength of $U(m)_k$
and $\mu , \nu = 0,1,2,3$,
while $g$ is the gauge coupling constant. 
We should read $A_{N+k}^\mu = A_{k}^\mu ,~etc$.  
$U_{k,k+1}$, called a link field, is transformed as
\begin{equation}
U_{k,k+1} \to L_k U_{k,k+1} L_{k+1}^\dagger ~, ~~~L_k \in U(m)_k,
\label{transformation-U}
\end{equation}
under $U(m)_k$.
The covariant derivative is defind as
$D^\mu U_{k,k+1} \equiv \partial^\mu U_{k,k+1} - i A^\mu_k U_{k,k+1} + i U_{k,k+1} A^\mu_{k+1}$. 

We may use a ``moose'' or ``quiver'' diagram to describe this theory. 
In such a diagram, gauge groups are represented by open circles,
and link fields by single directed lines attached to these circles.
Open circles and single directed lines are sometimes called sites and links.
The geometry built up from sites, links, and faces is sometimes called ``theory space''. 
These geometrical objects are identified as gauge groups, fields and potentials in the action.
The moose diagram characterizing the transformation (\ref{transformation-U}) is 
an $N$-sided polygon.

We assume that the absolute value of each link field
$|U_{k,k+1}|$ has the same value, $f$.
Then $U_{k,k+1}$ is expressed as
\begin{equation}
 U_{k,k+1} = f \exp (i {\chi} _{k}/f) ~.
\end{equation}
The $U_{k,k+1}$ kinetic terms go over to a mass-matrix for the gauge fields.
The gauge boson $(mass)^2$ matrix for $N=5$ is
\begin{equation}
g^2f^2
\left(\begin{array}{ccccc}
2      &    -1  &      0 &      0 &    -1  \\
-1     &     2  &     -1 &      0 &     0   \\
0      &    -1  &      2 &     -1 &     0   \\
0      &     0  &     -1 &     2  &    -1   \\
-1     &     0  &      0 &    -1  &     2  
\end{array}\right) ~~.
\label{gauge-boson_mass-matrix}
\end{equation}
We obtain the gauge boson mass spectrum:
\begin{equation}
M_{p}^{2} = 4g^2 f^2 \sin ^2 \left(\frac{\pi p}{N} \right) , ~~~~
p \in {\bf Z} ~, ~
\label{gauge-boson_mass-spectrum}
\end{equation}    
by diagonalizing (\ref{gauge-boson_mass-matrix}).

For $|p| \ll N$, the masses become 
\begin{equation}
M_p \simeq \frac{2\pi |p|}{r} ~,
\end{equation}
where $r \equiv Nb$ and $b \equiv 1/gf$.
This is precisely the Kaluza-Klein spectrum for a five-dimensional gauge boson
compactified on a circle of circumference $r$.

\subsection{Spectral Graph Theory}
In general, the theory space does not necessarily have a continuum limit. 
Sites can be complicatedly connected by links.
Such a connection is a {\it graph}.
We identify the theory space as a graph consisting of vertices and edges,
which correspond to sites and links, respectively.
Therefore,
DD can be generalized to field theory on a graph \cite{KSJMP}.

A graph $G$ consists of a {\it vertex} set $V(G) \neq \emptyset$
and  an {\it edge} set $E(G) \subseteq V(G) \times V(G)$, 
where an edge is an unordered pair of distinct vertices of $G$.
The {\it degree} of a vertex $v$, denoted by $deg(v)$, 
is the number of edges incident with $v$.

There are various matrices that are naturally associated with a graph.                                                      
The {\it graph Laplacian} (or {\it combinatorical Laplacian} ) 
$\Delta (G)$ is defined by \\
\begin{equation}
(\Delta)_{vv^{\prime}} 
      = \left\{\begin{array}{cl}
            deg(v) \quad&  {\rm if~}v = v^{\prime}           \\ 
             -1    \quad&  {\rm if~}v{\rm ~is~adjacent~to~}v^{\prime}  \\
            ~~0   \quad&  {\rm otherwise}
         \end{array}\right. ~ .
\end{equation}                        

For example, we consider a {\it cycle} graph,
which is equivalent to a moose diagram. 
The cycle graph with $p$ vertices is denoted by $C_p$.
For $C_5$, the Laplacian matrix takes the form:
\begin{equation}
\Delta(C_5)=
\left(\begin{array}{ccccc}
2  & -1 & 0  &0  &-1 \\
-1 & 2  & -1 &0  &0 \\
0  & -1 & 2  &-1 &0 \\
0  & 0  & -1 &2  &-1 \\
-1 & 0  &0   &-1 &2
\end{array}\right) \, .
\label{cm}
\end{equation}
Up to the dimensionful coefficient $g^2f^2$,
this matrix is identified with 
the gauge boson $(mass)^2$ matrix (\ref{gauge-boson_mass-matrix}).
We find, indeed, any theory space can be associated with the graph
and the $(mass)^2$ matrix for a field on a graph can be expressed by the graph Laplacian
owing to the Green's theorem for a graph.
                  
\section{Our story thus far}
We have constructed models of induced gravity by using several graphs \cite{KSPTP}.
With the help of knowledge of spectral graph theory,
we can easily find that the UV divergent terms concern the graph Laplacian in DD or theory on a graph.
Therefore, the UV divergences can be controlled by the graph Laplacian
and we can construct the models of one-loop finite induced gravity from a graph.

In the model \cite{KSPTP}, the one-loop finite Newton's constant is induced
and the positive-definite cosmological constant can also be obtained.

\section{Self-consistent Einstein Universe ($T\times S^3$)}
The metric of the static Einstein Universe \cite{EU}\cite{KKEU} is given by 
\begin{equation}
ds^2=-dt^2+a^2\left[d\chi^2+\sin^2\chi(d\theta^2+\sin^2\theta d\phi^2)\right],
\end{equation}
where $a$ is the scale factor
and $0 \le \chi \le \pi$, $0 \le \theta \le \pi$ and $0 \le \phi \le 2\pi$.
At finite temperature $T$, the one-loop effective action is regarded as free energy $F(a,\beta)$
and the Einstein equation becomes  
\begin{equation}
\frac{\partial(\beta F)}{\partial \beta}=\frac{\partial(\beta F)}{\partial a}=0,
\label{Einstein-equation}
\end{equation}
where $\beta\equiv 1/T$.
In this paper, we study self-consistent Einstein Universe in theory on a graph. 
In our models, four-dimensional fields are on cycle graphs.
The first model is that 
scalar fields are on 8 $C_{N/2}$, $U(1)$ vector fields on 4 $C_N$ 
and Dirac fermions on 2 $C_{N/2}$ + 3 $C_N$.
The second model is that 
scalar fields are on 16 $C_{N/4}$ + 2 $C_{N/2}$, vector fields on 5 $C_N$ 
and Dirac fermions on 4 $C_{N/4}$ + 3 $C_N$ + 2 $C_{N/2}$.
In each model,
Newton's constant and the cosmological constant are calculable 
and are not given by hand. 
%

\section{Results}
\begin{figure}
\begin{center}
\begin{minipage}{50mm}
\begin{center}
\includegraphics[height=5.2cm]{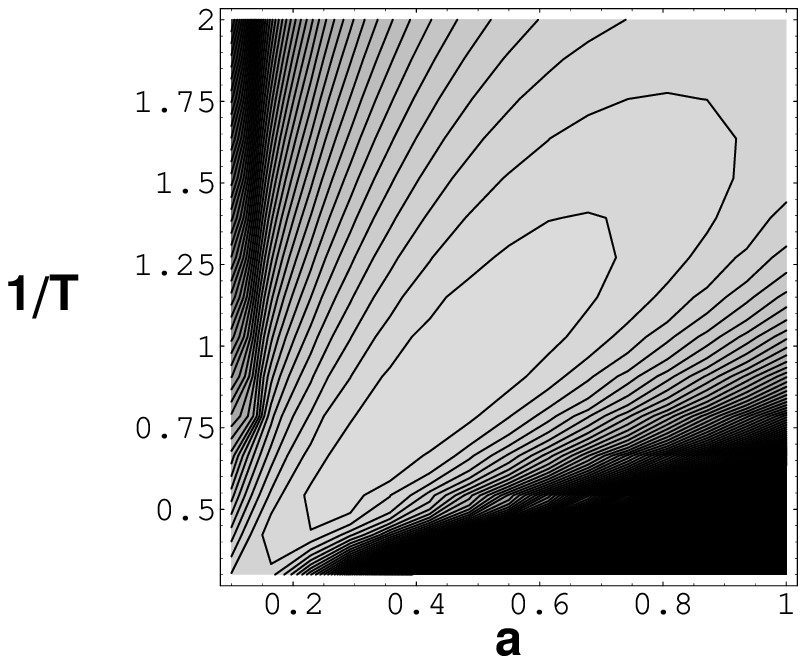}
\caption{
A contour plot of $\beta F$ in the first model, in which scalars on 8 $C_{N/2}$, vectors on 4 $C_N$ and 
Dirac fermions on 2 $C_{N/2}$+ 3 $C_N$.
A solution of the Einstein equation can be found at the maximum point.
}
\label{fig1}
\end{center}
\end{minipage}
\hspace{20mm}
\begin{minipage}{50mm}
\begin{center}
\includegraphics[height=5cm]{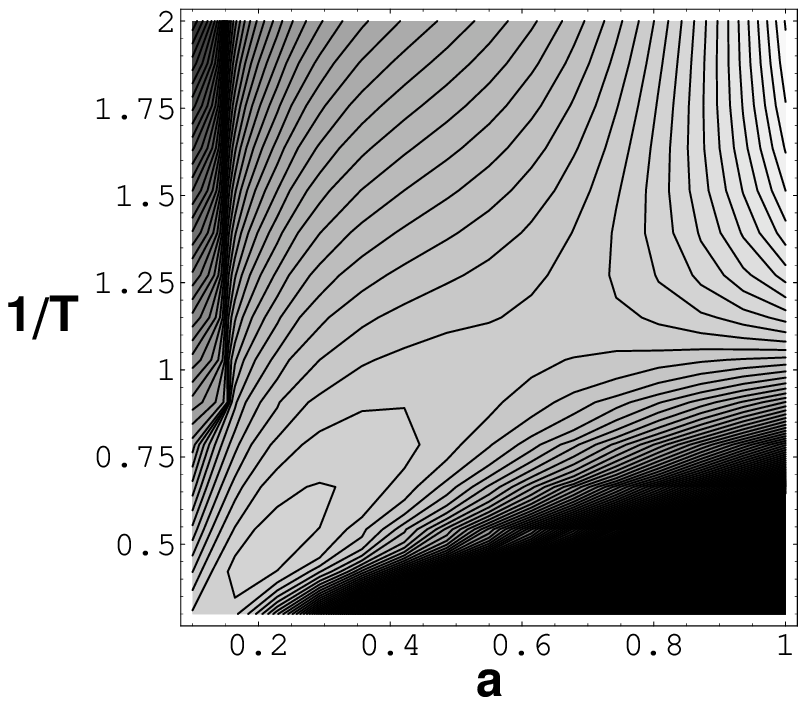}
\caption{
A contour plot of $\beta F$ in the second model, in which scalars on 16 $C_{N/4}$+2 $C_{N/2}$, 
vectors on 5$C_N$ and  Dirac fermions on 4 $C_{N/4}$ + 3$C_N$ + 2 $C_{N/2}$.
Two solutions of the Einstein equation can be found at the maximum and at the saddle point.
}
\label{fig2}
\end{center}
\end{minipage}
\end{center}
\end{figure}

We exhibit $\beta F$
for the first model in Fig.~\ref{fig1}
and for the second in Fig.~\ref{fig2}, for large $N$.
The horizontal axis indicates the scale factor $a$, 
while the vertical one indicates the inverse of temperature $T$. 
The scale of each axis is in the unit of $N/f$.
In the first model, the cosmological constant is zero 
and the solution 
can be found at the maximum of $\beta F$, 
corresponding to be in Casimir regime \cite{EU}.
In the second model,
the solution in Casimir regime and the solution in Planck regime \cite{EU} are found.

\section{Summary and Prospects}
We have studied self-consistent Einstein Universe in the graph theory space.
The solutions can be systematically obtained with the help of the graph structure.

As the future works,
we should investigate the possibility of obtaining the small cosmological constant 
and the large Plank scale 
in a model that scalar fields are on 4 $G_{(1)}$, vector fields on 4 $G_{(2)}$ and
Dirac fermions on $G_{(1)}$ + 3 $G_{(2)}$, while $\#V(G_{(1)})=\#V(G_{(2)}$).
We also should investigate the model with the time-dependent scale factor, $a(t)$.

In the present analysis, we have constructed models by using cycle graphs, 
but we are also interested in the model of general graphs.
For a $k$-regular graph,
the trace formula \cite{tf} is useful 
if we have a single mass scale.
Field theory on {\it weighted} graphs, which might correspond to warped spaces
in the continuous limit or not, is also interesting.
A quasi-continuous mass spectrum is conceivable
and
dynamics of graphs such as Hosotani mechanism is also thinkable. 
 
We expect that the knowledge of spectral graph theory produces useful results on deconstructed theories
and open up another possibilities of gravity models.

\section*{Acknowledgements}
I would like to thank T.~Hanada for useful comments, and also the organizers of JGRG17.

\end{document}